\documentclass[twocolumn,showpacs,preprintnumbers,amsmath,amssymb]{revtex4}
\usepackage{color}
\usepackage{comment}
\usepackage{amsmath}
\usepackage{amssymb}
\usepackage{amsthm}
\usepackage{amscd}
\usepackage{cases}
\usepackage{graphicx}
\usepackage{indentfirst}

\newcommand {\nn}    {\nonumber}
\newcommand{\pt}{\partial}

\newcommand{\be}{\begin{equation}}
\newcommand{\ee}{\end{equation}}
\newcommand{\ba}{\begin{eqnarray}}
\newcommand{\ea}{\end{eqnarray}}
\newcommand{\mcal}{\mathcal}
\begin{document}

\title{Brane worlds in critical gravity}

\author{Feng-Wei Chen$^1$\footnote{chenfw10@lzu.edu.cn}}
\author{Yu-Xiao Liu$^{1,2}$\footnote{liuyx@lzu.edu.cn}}
\author{Yuan Zhong$^1$\footnote{zhongy2009@lzu.edu.cn}}
\author{Yong-Qiang Wang$^1$\footnote{yqwang@lzu.edu.cn, corresponding author}}
\author{Shao-Feng Wu$^{3,4}$\footnote{sfwu@shu.edu.cn}}
 \affiliation{$^{1}$Institute of Theoretical Physics,
              Lanzhou University, Lanzhou 730000, China \\
          {$^2$Interdisciplinary Center for Theoretical Study,
University of Science and Technology of China, Hefei, Anhui 230026,
China}\\
{$^{3}$Enrico Fermi Institute, University of Chicago, Chicago, Illinois 60637, USA}\\
{$^{4}$Department of Physics, Shanghai University, Shanghai 200444, China}
}

\begin{abstract}
Recently, L\"u and Pope proposed critical gravities in [Phys. Rev. Lett. 106, 181302 (2011)]. In this paper we construct analytic brane solutions in critical gravity with matter. The Gibbons-Hawking surface term and junction condition are investigated, and the thin and thick brane solutions are obtained. All these branes are embedded in five-dimensional anti-de Sitter spacetimes. Our solutions are stable against scalar perturbations, and the zero modes of scalar perturbations {cannot} be localized on the branes.
\end{abstract}

\date{\today}






\maketitle

\section{Introduction}

It has been known that, by adding {higher-order derivative terms} (such as the squared-curvature terms) to the Einstein-Hilbert action, power-counting renormalizable theories of gravity can be realized. In the absence of the cosmological constant, although the theory is renormalizable, it suffers from having ghosts and is perturbatively nonunitary \cite{StellePRD1977,StelleGRG1978}.

Recently, motivated by the works of chiral topologically massive gravity with a negative cosmological constant in three dimensions \cite{Strominger,Bergshoeff}, critical gravities (quadratic-curvature actions with a cosmological constant) in four and higher dimensions have been constructed \cite{LuPRL2011,LuPRD2011a}.
At the critical point, these theories possess an {anti-de} Sitter (AdS) vacuum, for which there is only a massless tensor, and the linearized excitations have vanishing energy. It was also shown that at the critical point the theory admits additional modes, namely, the so-called logarithmic modes \cite{LuPRL2011,GrumillerJHEP2008a,AlishahihaPRD2011a,BergshoeffPRD2011a}, which arise as limits of the massive {spin-2} modes of the noncritical theory \cite{BergshoeffPRD2011a}. The quantization of the linear fluctuations of these critical gravities was studied in Ref.~\cite{PorratiPRD2011a}.
{The unitarity of critical gravity theories was} studied in Refs.~\cite{Gullu2011b,Ohta2011a}.


{It was shown} that critical gravity theories without matter fields in higher dimensions
admit solutions of the Einstein metrics ($R_{MN}=\Lambda g_{MN}$), which include both the AdS vacua and Schwarzschild-Tangherlini AdS black holes \cite{LuPRL2011,LuPRD2011a,PorratiPRD2011a,LiuLuLuo2011}. In Ref. \cite{GulluPRD2011a}, the authors found exact AdS-wave solutions in a general quadratic gravity theory with a cosmological constant.
It turns out that some of these solutions do affect the asymptotic structure of the AdS space via their logarithmic behavior.

However, vacua with constant curvatures appear only in special theoretical models. Most gravitational models study deviation from vacua. {Moreover}, some new properties of the critical gravity {appear} only in models with matter fields. So
it is crucial to find analytic background solutions.
In this paper, we focus on the Randall-Sundrum (RS) brane model, which offers us a solution to the hierarchy problem by embedding two 3-branes in an AdS$_5$ spacetime \cite{rsII,rsI}. In the original {setup}, gravity is described by {the Einstein gravity}.
There were some works about brane in higher derivative gravities
(see {for example} Refs. \cite{Nojiri:1999nd,Nojiri:2000gv,Giovannini:2001ta}).
Here {we would} like to reconstruct {a brane} model in the simplest higher derivative gravity but at the critical point
and give some exact solutions.
Although it is still not clear whether the critical gravity theory is renormalizable in the presence of matter, it is interesting to consider {a brane} model in this theory.
These considerations led us to the question: does critical gravity support RS brane solutions? {Also, how higher-order} curvature terms affect the properties of the solutions, for instance, the stability against linear perturbations, the junction conditions, etc?

In this paper, both the RS thin and thick branes with {codimension} one are considered. It is found that at the critical point the equations of motion are { of second order}, and brane solutions are found to be simple. For simplicity, we only investigate {Minkowski} branes, the generalization to AdS and dS branes will be considered in our future work.


\section{Junction condition and thin brane solutions in critical gravity}\label{sec2}

\subsection{The model}
First, we consider the thin brane in the five-dimensional critical gravity. The action is
\begin{eqnarray}
 S=S_{\text{g}} +S_{\text{b}},\label{actionS1}
\end{eqnarray}
where the gravity part $S_{\text{g}}$ and the brane part $S_{\text{b}}$ are given by
\begin{subequations}\label{SgSb}
\begin{eqnarray}
 S_{\text{g}} \!\!\!&=&\!\!\! \frac{1}{2\kappa^2}
    \int_M
     \Big[ R\!-3\Lambda_0
           \!+\! \alpha R^2
           \!+\! \beta R_{\!M\!N}\!R^{\!M\!N}
           \!+\! \gamma \mathcal{L}_{\text{GB}}
    \! \Big], ~~~~  \label{SG1}\\
 S_{\text{b}} \!\!\!&=&\!\!\! \int_{{\partial}M}~ (-V_0), \label{SB}
\end{eqnarray}\end{subequations}
where $ \int_M \equiv \int_M d^{5}x\sqrt{-g}$, $\int_{{\partial}M}\equiv \int d^{4}x\sqrt{-q}$,  $\mathcal{L}_{\text{GB}}=R^2-4R_{MN}R^{MN}+R_{MNPQ}R^{MNPQ}$ is the Gauss-Bonnet term, $q_{\mu\nu}$ is the induced metric on the brane, and $V_0$ is the brane tension. The capital Roman alphabets $M,N,...=0,1,2,3,4$ and the Greek letters $\mu,\nu,...=0,1,2,3$ denote the indices of the bulk and the brane, respectively.
The {line element} describing a static flat brane can be assumed as
\begin{equation}\label{line-element}
 ds^{2}=g_{MN}dx^M dx^N
  =\text{e}^{2A(y)}\eta_{\mu\nu}dx^{\mu}dx^{\nu}+dy^{2},
\end{equation}
where $\text{e}^{2A}$ is the warp factor with the normalized condition $\text{e}^{2A(0)}=1$. We introduce the $Z_2$ symmetry by setting $A(y)=A(-y)$.

The equations of motion are given by
\begin{eqnarray}
 \mathcal{G}_{MN}
     + \alpha E_{MN}^{(1)}
     + \beta E_{MN}^{(2)}
     -\frac{1}{2} \gamma H_{MN}
  = \kappa^2 T_{MN},
         \label{EOMThinBrane1}
\end{eqnarray}
where $T_{MN}=-V_0 \delta^{\mu}_{M} \delta^{\nu}_{N} g_{\mu\nu} \delta(y)$, and
\begin{eqnarray}
 \mathcal{G}_{MN} \!\!&=&\!\!  R_{MN}-\frac{1}{2} R~ g_{MN}
     + \frac{3}{2}\Lambda_0 g_{MN}, \nonumber \\
 E_{{MN}}^{(1)} \!\!&=&\!\! 2 R \Big(R_{{MN}}
       \!-\!\frac{1}{4}R ~g_{{MN}}\Big)
      \! +\! 2 g_{{MN}} \square R
      \! -\! 2\nabla_{\!M}\nabla_{\!N} R,  \nonumber\\
 E_{{MN}}^{(2)} \!\!&=&\!\!
         2 R^{{PQ}} \Big(R_{{MPNQ}}
       - \frac{1}{4} R_{{PQ}}~g_{{MN}}\Big) \nonumber \\
      &&\!\! + \square \Big(R_{{MN} }
       + \frac{1}{2}R~g_{{MN}}\Big)
       - \nabla _M\nabla_N R. \nonumber\\
 H_{MN}\!\!&=&\!\!g_{MN}\mathcal{L}_{\text{GB}}
         -4R R_{MN}
           +8R_{MP}R^P_N \nonumber \\
        &&\!\!  +8R_{MANB}R^{AB}
           -4R_{MABC}R_{N}^{~~ABC} .
\end{eqnarray}

The junction condition is determined by
\begin{eqnarray}
  \int_{0^-}^{0^+} dy \Big[\mathcal{G}_{\mu \nu }
     + \alpha E_{\mu \nu }^{(1)}
     + \beta E_{\mu\nu }^{(2)} -\frac{1}{2} \gamma H_{\mu\nu}\Big] \nonumber     \\
  =-\kappa^2 V_0 ~g_{\mu \nu }(0).
         \label{junctionConditions1}
\end{eqnarray}

It is very difficult to find thin brane solutions for arbitrary $\alpha$, $\beta$, and $\gamma$ for the fourth-order differential equations (\ref{EOMThinBrane1})  and the junction condition (\ref{junctionConditions1}). However, at the critical point $16\alpha +5\beta=0$ \cite{LuPRL2011,LuPRD2011a}, the equations of motion (EOMs) { in the bulk} are reduced to the following second-order ones:
\begin{subequations}\label{EOMThinBrane_c}
\begin{eqnarray}
  \Lambda_0+4 A'^2 + \zeta  A'^4 \!\!&=&\!\!0, ~~~~~\label{EOMThinBrane1c} \\
   \big(2+ \zeta A'^2\big) A''
    \!\!&=&\!\! 0,  \label{EOMThinBrane2c}
\end{eqnarray}
\end{subequations}
and the junction condition reads
\begin{eqnarray}
  \int_{0^-}^{0^+} dy
     \frac{3}{2} \big(2+\zeta A'^2\big) A''
     =\Big({3A'+\frac{\zeta}{2}A'^3}\Big)\Big|_{0^-}^{0^+}
     =-\kappa^2 V_0 ,
         \label{junctionConditions2}
\end{eqnarray}
where the prime denotes the derivative with respect to $y$, and
\begin{eqnarray}
  \zeta \equiv 3 \beta-8\gamma.     \label{zeta}
\end{eqnarray}
In the four-dimensional critical gravity, the square-curvature modifications have no effect on the brane solutions, and the Einstein equations are $\Lambda_0+3A'^2=0$ and $A''=0$.

\subsection{Junction condition}

Actually, for the general coefficients $\alpha$ and $\beta$ in a five-dimensional spacetime, we have the following identity:
\begin{equation}
  \alpha R^2+\beta R_{MN}R^{MN}+\gamma\mathcal{L}_{\text{GB}}=\frac{3\beta}{8}C^2
   -\frac{\zeta}{8}\mathcal{L}_{\text{GB}}+\frac{16\alpha+5\beta}{16}R^2.
\end{equation}
Here $C^2:=C^{MNPQ}C_{MNPQ}$ is the square of the five-dimensional Weyl tensor,
\begin{eqnarray}
  C_{MNPQ}&=&R_{MNPQ}+g_{MQ}S_{NP}-g_{NQ}S_{MP} \nonumber \\
          && + g_{NP}S_{MQ}-g_{MP}S_{NQ},\\
  S_{MN}&=&\frac{1}{3}\big(R_{MN}-\frac{1}{8}R g_{MN}\big). \label{SMN}
\end{eqnarray}
It is obvious that $16\alpha+5\beta=0$ and $\zeta=8\gamma-3\beta=0$ are special.
Since the Weyl tensor vanishes in our model, when the first condition is satisfied, i.e., $16\alpha+5\beta=0$, the solutions of the EOMs as well as the junction condition are the same as the Einstein-Gauss-Bonnet (EGB) gravity.

\subsubsection{Gibbons-Hawking method}

We can also adopt {the Gibbons-Hawking} method to derive the junction condition. First, we {outline the} basic idea. The thin brane divides the whole spacetime $M$ into two {submanifolds} and should be interpreted as the boundary $\partial M$ of the two {submanifolds}. $n^Q$ is the unit vector normal to the boundary $\partial M$ and outward pointing. $q^{MN}=g^{MN}-n^M n^N$ is the induced metric on the brane.  $K_{MN}=\mathcal{L}_{\vec{n}} q_{MN}/2 $ is the extrinsic curvature ($\mathcal{L}_{\vec{n}}$ denotes { the Lie} derivative in the direction $\vec{n}$), and $K=g^{MN}K_{MN}$ is the trace of the extrinsic curvature. An important property is that the directions of $n^Q$ on both sides of $\partial M$ are opposite. If we fix the vector $n^Q$, the final results can be written as $[\bullet]_\pm$, where $[F]_{\pm}:=F(0+)-F(0-)$\footnote{It means when we calculate $F(0\pm)$, $n^Q$ is kept fixed, not $n^Q(0\pm)$, i.e., for example, $K_{MN}=-\pt_y q_{MN}/2$ for both sides.}. See e.g. Refs. \cite{ParryPichlerDeeg2005,Balcerzak:2007da} for the details. In the following,  we choose $n^{Q}(0+)=n^{Q}:=(0,0,0,0,-1)$ for right side, and only calculate right side.

The Gibbons-Hawking surface term
of the EGB theory was given in Refs. \cite{Deruelle:2000ge, Davis:2002gn, Maeda:2003vq}:
\begin{eqnarray}
S_{\text{EGB-surf}}=\frac{1}{\kappa^2}\int_{\partial M}
  \Big(K-\frac{\zeta}{4}(J-2\tilde{G}_{\mu\nu}K^{\mu\nu})\Big).
\end{eqnarray}
Here $\tilde{G}_{\mu\nu}=\tilde{R}_{\mu\nu}-q_{\mu\nu}\tilde{R}/2$ is the Einstein tensor of the induced metric $q_{\mu\nu}$ and $J$ is the trace of the following tensor:
 \begin{eqnarray}
   J_{MN}&=&\frac{1}{3}\Big(2K K_M^P K_{PN}+K^{PQ}K_{PQ}K_{MN} \nonumber \\
   &&-K^2K_{MN}-2K_{MP}K^{PQ}K_{QN}\Big).
 \end{eqnarray}

The junction condition for the EGB theory is (in the following, we will prove that the contribution from the $C^2$ term vanishes for the conformally flat case)
\begin{eqnarray}
 E_{\text{GB}}^{\mu\nu}&:=& [K_{\mu\nu}]_{\pm}-q_{\mu\nu}[K]_{\pm} \nonumber \\
      &&-\frac{\zeta}{4}
      \Big(3[J_{\mu\nu}]_{\pm}
       -q_{\mu\nu}[J]_{\pm}
       -2P_{\mu\rho\nu\sigma}[K^{\rho\sigma}]_{\pm}\Big)\nonumber \\
     &=& -\kappa^2 V_0 q_{\mu\nu}(0),\label{EGB JunctionCondition}
\end{eqnarray}
where
\begin{equation}
   P_{\mu\nu\rho\sigma}
     = \tilde{R}_{\mu\nu\rho\sigma}
      +2q_{\mu[\sigma}\tilde{R}_{\rho]\nu}
      +2q_{\nu[\sigma}\tilde{R}_{\rho]\mu}
      +\tilde{R}q_{\mu[\rho}q_{\sigma]\nu}.
 \end{equation}
In our case, $q_{\mu\nu}=\eta_{\mu\nu}e^{2A(y)}$, $A(0)=0$ and $K_{\mu\nu}(0+)=-K_{\mu\nu}(0-)=-A'(0+)\eta_{\mu\nu}$. Eq. (\ref{EGB JunctionCondition}) gives the same result of Eq. (\ref{junctionConditions2}).

For a general warped geometry with $ds^2=e^{2A(y)}\hat g_{\mu\nu}(x)dx^\mu dx^\nu+dy^2=e^{2A}(\hat g_{\mu\nu}dx^\mu dx^\nu+dz^2)$, the junction condition is also {of first order} in the critical gravity, because $C^{MPNQ}$ is continuous. However, in this case, the solutions of the EGB gravity do not satisfy the EOMs of the critical gravity. {We can also prove this statement from the full variational principle. This means that we should start from the action of the general case instead of the warped geometry. That will be more convincing.} Explicitly, we have
\begin{eqnarray}
\!\!\!\!\!\!   \delta\int_M  C^2 
  &=& \int_M \Big[2C_{M}^{~PQR}C_{NPQR}-\frac{1}{2}g_{MN}C^2 \nonumber \\
  &+& \frac{8}{3}R^{PQ}C_{MPNQ}-4C^{~~~P~~Q}_{(M~N)~;PQ}\Big]
       \delta g^{MN}\nonumber\\
  &+&   4\int_{\partial M} \Big[(C^{MPNQ}n_Q \delta g_{MN})_{;P} \nonumber \\
  &-&  \left((C^{MPNQ}n_Q)_{;P}+C^{MPNQ}_{~~~~~~;Q}n_P \right) \delta g_{MN}\Big]. \label{Cboundary2}
\end{eqnarray}

The bulk term gives contribution to the EOMs, and the boundary term (and the corresponding generalized Gibbons-Hawking term) will give contribution to the corresponding junction condition.

In order to have a well-posed variational principal, we introduce an auxiliary field $\varphi^{MNPQ}$, which has the same symmetry as the Weyl tensor and is also totally traceless. So $C^2$ is replaced by $2\varphi^{MNPQ}C_{MNPQ}-\varphi^{MNPQ}\varphi_{MNPQ}$. Its EOM is $\varphi^{MNPQ}=C^{MNPQ}$. Then we replace $C^{MNPQ}$ by the new field $\varphi^{MNPQ}$ in Eq. (\ref{Cboundary2}).

{To proceed}, we give some useful identities (for our case $a_N:=n^M n_{N;M}=0$):
 \begin{eqnarray}
  \delta n_M&=&-\frac{1}{2}n_M n_P n_Q\delta g^{PQ},\label{basic1}\\
 X^M_{;M}&=&D_M(q^{M}_N X^N)+ K n_NX^N+\mathcal{L}_{\vec{n}}(n_NX^N).\label{basic2}
 \end{eqnarray}
Since $n_M$ is the unit norm to a hypersurface (brane), we obtain $n_M=\frac{\partial_M T}{\sqrt{g^{PQ}\partial_P T \partial_Q T}}$ for some function $T(x,y)$. This {will} lead to the identity (\ref{basic1}) immediately.
The second one can be proven straightforward:
\ba
D_M(q^N_P X^P):&=&q_M^Q q^N_R(q^R_P X^P)_{;Q},\\
 D_M(q^M_P X^P)&=&q_M^Q q^M_R(q^R_P X^P)_{;Q}
                 =q^Q_R(q^R_P X^P)_{;Q}\nn\\
               &=& q^Q_RX^R_{;Q}-q_M^Nn^M_{;N}n_Q X^Q\nn\\
               &=&X^Q_{;Q}-(n^Q n_RX^R_{;Q}+K n_QX^Q) \nn\\
               &=&X^Q_{;Q}-\mathcal{L}_{\vec{n}}( n_QX^Q)-K n_QX^Q.
\ea
With the help of the identity (\ref{basic2}), after integrating out the pure divergence $D_M(q^{M}_N X^N)$, we have
\begin{eqnarray}
  &&4\int_{\partial M} (\varphi^{MPNQ}n_Q \delta g_{MN})_{;P}\nonumber\\
\!\!\!\!\!\!\!\! &=&4\int_{\partial M} \Big[
     \varphi^{MPNQ}n_Q n_P\mathcal{L}_{\vec{n}} \delta g_{MN}  \nonumber\\
  &+& \Big(K\varphi^{MPNQ}n_Q n_P
          +\mathcal{L}_{\vec{n}}(\varphi^{MPNQ}n_Q n_P)
     \Big) \delta g_{MN}
    \Big]. \label{phinndg}
\end{eqnarray}
We define a new tensor $\varphi^{MN}:=\varphi^{MPNQ}n_Q n_P$, which has the properties: $\varphi^{MN}=\varphi^{NM}$, $\varphi^{MN}n_N=0$ and $\varphi^{MN}g_{MN}=\varphi^{MN}q_{MN}=0$. Then, the first term in Eq. (\ref{phinndg}) gives
\begin{eqnarray}
  && \varphi^{MN}\mathcal{L}_{\vec{n}} \delta g_{MN} \nonumber\\
 &=&\varphi^{MN}\left[\delta(\mathcal{L}_{\vec{n}}g_{MN})-2g_{PM}(\delta n^P)_{;N}\right]\nonumber\\
 &=&\varphi^{MN}\left[\delta(\mathcal{L}_{\vec{n}}g_{MN})+n_{N;M}n_P n_Q\delta g^{PQ}+2(n^P\delta g_{PM})_{;N}\right]\nonumber\\
 &=&\varphi^{MN}(2\delta K_{MN}-K_{MN}n^P n^Q\delta g_{PQ})+2\varphi^{MN}(n^P\delta g_{PM})_{;N}\nonumber\\
 &=& 2\varphi^{MN}\delta K_{MN}
     -\varphi^{MN}K_{MN}n^P n^Q\delta g_{PQ}\nonumber\\
 &&  +2D_N(\varphi^{MN}n^P\delta g_{PM})
     -2\varphi^{PM}_{;P}n^N\delta g_{MN}.
\end{eqnarray}
So the surface term for the $3{\beta}C^2/8$ part is
\begin{eqnarray}
\!\!\!\!\!  \delta S_{C^2} &=& \delta \int_M  \frac{3\beta}{8} S_{C^2} \nonumber \\
 &=& \frac{3\beta}{4\kappa^2}\int_{\partial M}
      \Big\{ 2\varphi^{MN}\delta K_{MN}
      +\Big[
        \mathcal{L}_{\vec{n}}\varphi^{MN}
       \nonumber\\
     && + K\varphi^{MN}
        - \varphi^{PQ}K_{PQ}n^M n^N
        - 2\varphi^{P(M}_{;P}n^{N)}        \nonumber\\
     &&
        - (\varphi^{MPNQ}n_Q)_{;P}
        - \varphi^{MPNQ}_{~~~~~~;Q}n_P \Big] \delta g_{MN}\Big\}. \label{c-bound-g1}
\end{eqnarray}
Then with $2\varphi^{MN}\delta K_{MN} = 2\varphi^{MN}\delta (K_{MN}-\frac{1}{4}q_{MN}K)+\frac{1}{2}K\varphi^{MN}\delta q_{MN}$, we get
\begin{eqnarray}
  \delta S_{C^2}
     &=& \frac{3\beta}{4\kappa^2}\int_{\partial M}
      \Big[ 2\varphi^{MN}\delta\bar{K}_{MN}  \nonumber\\
     &&    +\big(W^{MN}-\varphi^{PQ}K_{PQ}n^M n^N \big) \delta g_{MN}
      \Big],  \label{c-bound-g}
 \end{eqnarray}
 where
 \begin{eqnarray}
 \bar{K}_{MN}&=&K_{MN}-\frac{1}{4}q_{MN}K, \\
   W^{MN}&=&\frac{3}{2}K\varphi^{MN}+\mathcal{L}_{\vec{n}}\varphi^{MN}
       -2\varphi^{P(M}_{;P}n^{N)}  \nonumber \\
    && {-(\varphi^{MPNQ}n_Q)_{;P}
       -\varphi^{MPNQ}_{~~~~~~;Q}n_P}.
 \end{eqnarray}
It is not difficult to check the following identities:
\begin{equation}
W^{MN}n_M=0,~~~~W^{MN}q_{MN}=W^{MN}g_{MN}=0.
\end{equation}

Now we can introduce the corresponding Gibbons-Hawking surface term \cite{Hawking:NP1984} for the $C^2$ term
\begin{eqnarray}
  S_{\text{CGH}}=\frac{3\beta}{2\kappa^2}\int_{\partial M}~\varphi^{MN} \bar{K}_{MN}.
  \end{eqnarray}
So we have (considering the whole spacetime)
\begin{eqnarray}
\!\!\!\!  \delta (S_{C^2}+S_{\text{CGH}}) 
  &=& \frac{3\beta}{4\kappa^2}\int_{\partial M}
       \Big\{2 \big[\bar{K}_{MN}\big]_{\pm} {\delta \varphi^{MN}}  \nonumber \\
  &-&  \big[\varphi^{PQ}K_{PQ}\big]_\pm n^M n^N\delta g_{MN}\nonumber\\
  &+&  \big[W^{MN} -2\varphi^{P(M}\bar{K}^{N)}_P\big]_\pm\delta g_{MN} \Big\}.
\end{eqnarray}
The junction conditions are
 \begin{eqnarray}
 \!\!\!\!\! &&\big[\bar{K}_{MN}\big]_\pm =0,\label{barK}\\
 \!\!\!\!\! &&\big[\varphi^{PQ}K_{PQ}\big]_\pm=\bar{K}_{PQ}[\varphi^{PQ}]_\pm=0,\\
 \!\!\!\!\! &&-\frac{3\beta}{2}\big[W^{MN}-2\varphi^{P(M}\bar{K}^{N)}_P\big]_\pm
    +\big[E^{MN}_{\text{GB}}\big]_\pm
    =\kappa^2T_{(\text{brane})}^{MN}\label{c-jun}.
 \end{eqnarray}
Here $T_{(\text{brane})}^{MN}$ only contains the singular part of $T^{MN}$. We have omitted the continuous terms $q^{MN}\varphi^{PQ}K_{PQ}$ in Eq. (\ref{c-jun}). To avoid $\delta$-function in the junction conditions, we need stronger condition $[\varphi^{MN}]_\pm=0$ (like the {constraint} $[g_{MN}]_\pm=0$). Then it is easy to prove that the results {do not} depend on the choice of any basic field. For this case, Eq. (\ref{c-jun}) becomes
\ba
-\frac{3\beta}{2}\big[W^{MN}\big]_\pm+\big[E_{\text{GB}}^{MN}\big]_\pm =-\kappa^2T_{(\text{brane})}^{MN}\label{c-jun1}.
\ea

 Obviously, Eq. (\ref{barK}) gives no more {constraint} for brane solutions since $\bar{K}_{MN}\equiv0$.  {Also}
 the $C^2$ term {does not contribute} for the conformally flat spacetime.

\subsubsection{Another auxiliary field method}

There is another auxiliary field method that is widely used for critical gravity theories. Next we consider this method. The lagrangian (\ref{SG1}) can be written as
\begin{eqnarray}
  2\kappa^2\mathcal{L}
  &=& R-3\Lambda_0+\gamma\mcal{L}_{GB}
      +f^{MN}G_{MN} \nonumber \\
  &&  -\frac{1}{4\beta}(f_{MN}f^{MN}-f^2),\label{action-vac}
\end{eqnarray}
where the auxiliary field $f_{MN}$ is a symmetric tensor, and $f=f_{MN}g^{MN}$.
The EOM of the auxiliary field $f_{MN}$ is $f_{MN}=2\beta S_{MN}$ with $S_{MM}$ defined in Eq.~(\ref{SMN}).

From the lagrangian (\ref{action-vac}), we have (ignoring the EOM part and Gauss-Bonnet boundary part)
\ba
\delta(2\kappa^2\mathcal{L})=(B^{MNPQ}\delta g_{MN;P})_{;Q}-(B^{MNPQ}_{~~~~~~;Q}\delta g_{MN})_{;P}.
\ea
Here we have defined
\ba
F^{MN}   \!\!\! &\,=& \!\!\! f^{MN}+g^{MN}(1-\frac{1}{2}f), \\
 \!\!\!  B^{MNPQ} \!\!\! &:=&  \!\!\! F^{P(M}g^{N)Q}-\frac{1}{2}(g^{MN}F^{PQ}+g^{PQ}F^{MN}),~~~\\
B^{MN}   \!\!\! &:=&  \!\!\! B^{MNPQ}n_P n_Q.~~~(B^{MN}n_N=0.)
\ea
The field $B^{MN}$ plays {a similar role to} the field $C^{MN}$ except that $B^{MN}$ is not traceless.
Repeating the {steps} (\ref{Cboundary2})-(\ref{c-bound-g1}), we have
\ba
\delta S_{\text{g}}= \frac{1}{2\kappa^2}\int_{\partial M}
      \left( 2B^{MN}\delta K_{MN} +\Omega^{MN} \delta g_{MN}\right),
\ea
where
\ba
\Omega^{MN}&:=&KB^{MN}+\mathcal{L}_{\vec{n}}B^{MN}
      -B^{PQ}K_{PQ}n^M n^N
      \nonumber \\
   &+& 2B^{P(M}_{;P}n^{N)}
      +(B^{MPNQ}n_Q)_{;P}
      +B^{MPNQ}_{~~~~~~;Q}n_P.
\ea
The generalized Gibbons-Hawking term is
\ba
\!\!\! S_{\text{gGH}} \!\!\!&=&\!\!\! -\frac{1}{\kappa^2}\int_{\pt M} B^{MN}K_{MN}\nn\\
      &=&\!\!\! \frac{1}{2\kappa^2}\int_{\pt M}
         \Big(f^{MN}+2q^{MN}
              -q^{MN}f^{PQ}q_{PQ}
         \Big) K_{MN}.
\ea
The variation of the full action gives (the bulk and boundary terms of the Gauss-Bonnet term are omitted)
\ba
 \!\!\!&& {2\kappa^2} \delta (S_{\text{g}}+S_{\text{gGH}}) \nonumber \\
 \!\!\!&=& 
      \int_{\partial M} \Big[
          2 K_{MN}\delta B^{MN}
          +(B^{PQ} K_{PQ} q^{MN}
          +\Omega^{MN} )\delta g_{MN}\Big] \nonumber \\
 \!\!\!    &=& 
      \int_{\partial M}\Big[
          2 (K q_{MN}-K_{MN})\delta f^{MN}
          +\big(2K^{MN}\hat{f}
          -2K\hat{f}^{MN}\nn\\
 \!\!\!     &&  +B^{PQ} K_{PQ}q^{MN}-2K^{MN}+\Omega^{MN}\big) \delta g_{MN}\Big],\label{f-junction}
\ea
where $\hat{f}^{MN}=f^{PQ}q_P^M q_Q^N$ and $\hat{f}=q_{MN}\hat{f}^{MN}$.{It should be emphasized that we do not assume any ansatz of the background metric in the variational process. So it is also true for the general case.} It is suggested in Ref. \cite{Hohm:2010jc} that we can set the variation of the basic (or bare) field $\delta f_{MN}$ (or $\delta f^{MN}$) {to zero} on the boundary. However, the junction condition {depends} on the choice of the basic field. If we choose $f_{MN}$ as {the} basic field, using $\delta f^{MN}=g^{PM}g^{NQ}\delta f_{PQ}+2f^{(M}_P\delta g^{N)P}$, it { will give a different} junction condition unless $[(K q_{MP}-K_{MP})f^P_N]_\pm=0$ (this cannot be satisfied for our case).
What is worse, {neither of them give} consistent results. (The corresponding two Gibbons-Hawking terms are not the same, either.)

In our opinion, we cannot make the above assumption from the perspective of the variational principle, at least for the {higher-dimensional} critical gravity. Using the Gauss-Codazzi equation, we find that the {higher-order derivative term} in $f_{MN}$ only includes $\mathcal{L}_{\vec{n}}\bar{K}_{MN}$, and {the other part} should be dealt with as is done in {the Gauss-Bonnet gravity}.

In order {to obtain the} correct junction condition, irreducible components are very important. Taking $f(R)$ theories for example, the junction condition only requires $[K]_\pm=0$~\cite{Balcerzak:2007da}. For the (higher-dimensional) critical gravity, apart from the second-order EGB part, the action only includes the $C^2$ term. $C_{MNPQ}$ is an irreducible component {of the Riemann curvature, which results that the corresponding junction condition just contains the tensor $\bar{K}_{MN}$}.


\subsection{Thin brane solutions}

For $\zeta=0$ (i.e., $\gamma=3 \beta/8$), according to Refs. \cite{Nojiri:1999nd, Nojiri:2000gv}, the theory dual to the $\mathcal{N}=2$ superconformal field theory
is presumably related with the type IIB string on AdS$_5 \times X_5$, where $X_5=S_5/\mathbb{Z}_2$. The solution is {just a flat brane} in the Einstein gravity. It is also true for the AdS and dS branes. We do not give the solution here anymore. However, the linear fluctuation equations {in the critical gravity} are very different from those in { the Einstein gravity}.


In the following, we will give the solutions of the above brane equations
(\ref{EOMThinBrane_c}) for $\zeta\neq0$ (i.e., $\gamma\neq 3\beta/8$).

For $\zeta\neq0$, Eqs. (\ref{EOMThinBrane_c}) support two solutions:
\begin{subequations}
\begin{eqnarray}
 A_{\pm}(y)\!\!\! &=& \!\!\!
    -\sqrt{\frac{\pm \sqrt{4-\zeta\Lambda_0}-2}
                   {\zeta}
             }
    ~|y|,  \label{Aypm} \\
 V_{0\pm}~\!\!\! &=&  \!\!\! \frac{4\pm\sqrt{4-\zeta \Lambda_0}}{\kappa ^2}
                 \sqrt{\frac{\pm \sqrt{4-\zeta\Lambda_0}-2}
                   {\zeta}
             } , \label{V0pm}
\end{eqnarray}\label{ThinBraneSolutionB}
\end{subequations}
where the brane tensions are calculated with the junction condition (\ref{junctionConditions2}) or (\ref{EGB JunctionCondition}) or ({\ref{c-jun1}).

For the first brane solution, $A_{+}(y)$ and $V_{0+}$,  the {constraints} for the parameters are $\zeta>0$ and $\Lambda_0<0$, or $\zeta<0$ and $4/\zeta\leq\Lambda_0<0$. For both {constraints} the brane tension is positive.

For the second brane solution, $A_{-}(y)$ and $V_{0-}$, the {constraints} are $\zeta<0$ and $\Lambda_0\geq4/\zeta$.
The brane tension is positive  and negative for $4/\zeta \leq\Lambda_0<-12/\zeta$ and $\Lambda_0 > -12/\zeta$, respectively.
So, for this solution, the naked cosmological constant {can be vanishing}, for which we get a positive tension brane with the brane tension given by $V_{0-}=\frac{4}{\kappa ^2 \sqrt{-{\zeta}}}$.
Furthermore, it is interesting to note that when $ {\Lambda_0}=-{12}/{\zeta}$,
the brane tension $V_{0-}$ in (\ref{V0pm}) vanishes and the warp factor reduces to
$ A_{-}(y) = -\sqrt{\frac{\Lambda_0}{2}} ~|y|$.
Note that although the naked brane tension in the special case is zero, we could identify
$-\alpha E_{\mu\nu}^{(1)} -\beta E_{\mu\nu }^{(2)} +\frac{1}{2} \gamma H_{\mu\nu} $ as an effective energy-momentum term $\kappa^2 T_{\mu\nu}^{(\text{eff})}$ to get an effective positive brane tension.

From the two solutions (\ref{Aypm}), we have
$R_{MN}= -4{\frac{\pm \sqrt{4-\zeta\Lambda_0}-2} {\zeta}} g_{MN}=\Lambda g_{MN}$.
The effective cosmological constant $\Lambda$ is always negative, { irregardless of the sign of the naked cosmological constant $\Lambda_0$}. Therefore, the thin branes are embedded in five-dimensional AdS spacetimes with the cosmological constants $\Lambda=-4{\frac{\pm \sqrt{4-\zeta\Lambda_0}-2} {\zeta}}(<0)$.

Now, we study the limits of the solutions (\ref{ThinBraneSolutionB}) under $\zeta\rightarrow 0$. For the brane solution $A_{-}(y)$, the limit is divergent. While, for $A_{+}(y)$ and $V_{0+}$, they can be expanded as
\begin{eqnarray}
 A_{+}(y) \!\!\! &=& \!\!\! -\frac{1}{2}\sqrt{-\Lambda_0}
     \left( 1+\frac{\Lambda_0}{32}\zeta
     +\mathcal{O}(\zeta^2)   \right)|y|,  \label{Ay+}  \\
 V_{0+}~ \!\!\! &=& \!\!\! \frac{3}{\kappa^2}\sqrt{-\Lambda_0}
        \left( 1 -\frac{3\Lambda_0}{32}\zeta+\mathcal{O}(\zeta^2) \right) .  \label{V0+}
\end{eqnarray}
So, when $\zeta\rightarrow0$, the first brane solution in (\ref{ThinBraneSolutionB}) can be reduced to the RS brane solution, while the second one cannot.

At last, we mention that, when
\begin{eqnarray}
 \zeta =4/\Lambda_0, ~(\Lambda_0<0),  \label{zetaLambda0}
\end{eqnarray}
both solutions in Eq. (\ref{ThinBraneSolutionB}) become the same one:
\begin{eqnarray}
 A(y)&=&-\sqrt{-\Lambda_0/2}|y|,  \label{Ay0} \\
 V_0&=&2\sqrt{-2\Lambda_0}\kappa^{-2},
\end{eqnarray}
for which the effective cosmological constant also becomes the same one $\Lambda=2{\Lambda_0}$.

\section{Thick brane solution in critical gravity}\label{sec3}

Next, we consider the thick brane generated by a scalar field in the five-dimensional critical gravity. The action reads
\begin{equation}\label{action}
 S=S_{\text{g}}+S_{\text{m}},
\end{equation}
where $S_{\text{g}}$ is given by (\ref{SG1}) and the matter part is
\begin{equation}\label{action}
 S_{\text{m}}=\int_M 
     \Big(
      -\frac{1}{2}g^{MN}\partial_{M}\phi\partial_{N}\phi  -V(\phi)\Big). \nonumber
\end{equation}
The naked cosmological constant $\Lambda_0$ can be absorbed into the scalar potential.
The {line element} is also assumed as (\ref{line-element})
and the scalar field $\phi=\phi(y)$ for a static brane.

The EOMs for general $\alpha$ and $\beta$ are of fourth order, while they reduce to the following second-order ones at the critical point $16\alpha +5\beta=0$:
\begin{subequations}\label{EOMThickBrane}
\begin{eqnarray}
 -\frac{3}{2} \Big(\zeta A'^2 + 2 \Big) A''
   \!\!&=& \kappa^2 \phi'^2, \label{EOM1c} \\
 \frac{3}{2}\Big(\zeta A'^4\!+4 A'^2\!+\!\Lambda_0\Big)
   \!\!&=&\!\!\kappa^2 \Big(\frac{1}{2}\phi'^2\!-\!V\Big),~~~ \label{EOM2c} \\
 \phi'' \!+\! 4 A'\phi' \!\!&=&\!\! V_{\phi},\label{EOM3c}
\end{eqnarray}
\end{subequations}
where $V_{\phi}\equiv \frac{dV}{d\phi}$.
Note that Eq. (\ref{EOM3c}) can be derived from Eqs. (\ref{EOM1c}) and (\ref{EOM2c}). Hence, the above three equations are not independent.

In order to solve the above second-order differential equations, we can use the superpotential method. Introducing the superpotential function $W(\phi)$, the EOMs (\ref{EOM1c})-(\ref{EOM3c}) can be solved by the first-order equations:
\begin{subequations}\label{EOMThickBraneSuperpotential}
\begin{eqnarray}
 A' \!\!&=&\!\! -\frac{\kappa^2}{3} W,
            \label{EOM1csuperpotential} \\
 \phi' \!\!&=&\!\! \left(1+c_1 W^2\right)
            W_{\phi},\label{EOM2csuperpotential} \\
 V \!\!&=&\!\! \frac{1}{2}\left(1+c_1 W^2\right)^2  W_{\phi }^2
           -c_2 W^4  -c_3 W^2
           \!-\!\frac{3\Lambda_0}{2\kappa^2}, ~~~~\label{EOM3csuperpotential}
\end{eqnarray}
\end{subequations}
where $c_1=\frac{1}{18}\zeta\kappa^4$,
$c_2=\frac{1}{54}\zeta\kappa^6$,
and $c_3=\frac{2}{3}\kappa^2$.
Again, the parameters $\beta$ and $\gamma$ have no effect on the Einstein equations in four dimensions.

The energy density $\rho(y)$ of the system is given by
$\rho(y)=T_{MN}U^M U^N=-{T^{0}}_{0} =  \frac{1}{2}\phi'^2+V$.
For a brane solution, we require that the energy density on the boundaries of the extra dimension vanishes:
\begin{eqnarray}
 \rho(|y|\rightarrow \infty) \rightarrow 0, \label{BraneCondition}
\end{eqnarray}
from which the naked cosmological constant $\Lambda_0$ will be determined.

Next, we will give the solutions of the equations (\ref{EOMThickBraneSuperpotential}) with some choices of the superpotential.
When $\zeta=0$, these equations will reduce to the case of general relativity, which has been discussed widely. So we only consider the nontrivial case of $\zeta\neq0$, for which we can get the usual $\phi^4$ potential by setting $W=3a\phi$. The scalar potential is
\begin{equation}\label{Vphi_zeta}
 V(\phi)=b(\phi^2-v_0^2)^2,
\end{equation}
where
\begin{eqnarray}
 b &=&\frac{3}{8}
      \big(3 a^2\kappa^2\zeta^2-4\zeta \big)
      a^4\kappa^6, \nonumber \\
 v_0^2 &=& -\frac{2}{a^2\kappa^4\zeta},  \nonumber 
\end{eqnarray}
and the corresponding naked
cosmological constant is
\begin{eqnarray}
 \Lambda_0 = \frac{4} {\zeta}, \label{zeta_2}
\end{eqnarray}

When $\zeta>0$, the above scalar potential (\ref{Vphi_zeta}) is not a usual $\phi^4$ potential with two degenerate vacua since $v_0^2<0$. Such potential does not support a thick brane solution because the energy density is divergent at the boundaries of the extra dimension $y$.

So we are only interested in the case of $\zeta<0$, for which $v_0^2 >0$, $b>0$, and the above scalar potential (\ref{Vphi_zeta}) has two vacua at $\phi_{\pm}=\pm v_0$. The solution is
\begin{subequations}\label{{ThickBraneSolution_n}}
\begin{eqnarray}
 \phi(y)&=& v_0 \tanh(ky) , ~~~~(\zeta<0)\label{ThickBraneSolution_n_phi}\\
 e^{2A(y)}&=& \big[\cosh(ky)\big]
             ^{-\frac{2}{3} \kappa^2 v_0^2 } ,
             \label{ThickBraneSolution_n_E2A}
\end{eqnarray}
\end{subequations}
where $k=3a/v_0=3 a^2 \kappa^2\sqrt{-\zeta/2}$. This solution stands for a thick flat brane with the energy density given by
\begin{eqnarray}
 \rho(y)=
   \frac{1}{2} v_0^2
        \left(k^2+2b v_0^2\right)
  \text{sech}^4(ky).
\end{eqnarray}
The thickness of the brane is of about $1/k$.
On the boundaries $|y|\rightarrow\infty$, the solution of the warp factor is
\begin{eqnarray}
 A(|y|\rightarrow\infty) \rightarrow -\sqrt{\frac{-\Lambda_0}{2}} ~|y|.  \label{Ay_y_infinity}
\end{eqnarray}
Note that the asymptotic solution (\ref{Ay_y_infinity})
with the relation (\ref{zeta_2}) is in accord with the thin brane solution
(\ref{zetaLambda0})-(\ref{Ay0}) given in the previous section.
From the asymptotic solution (\ref{Ay_y_infinity}), we have $R_{MN}(|y|\rightarrow\infty) \rightarrow 2\Lambda_0 g_{MN}=\Lambda g_{MN}$. Therefore, the thick flat brane is embedded in an AdS spacetime with the cosmological constant $\Lambda=2\Lambda_0$.
In a general quadratic curvature gravity theory in $n(>4)$ dimensions without matter fields,
there are two disconnect AdS vacua. In the $n(>4)$-dimensional critical gravity, there is a unique critical vacuum \cite{LuPRD2011a}.

Note that in the above discussions we worked with double-well potentials with two ordered vacua and the domain walls interpolate between the two ordered vacua. An interesting {question is whether} one can also construct domain walls interpolating between the vacuum with $\phi=0$ and the ordered vacua. To this end, we {need to analysis} Eq. (\ref{EOMThickBraneSuperpotential}). At the boundaries $y\rightarrow\pm\infty$, we set $\phi(-\infty)=v$, $\phi(+\infty)=0$, and $A(y\rightarrow\pm\infty)\rightarrow-k|y|$ with $k>0$, and so $\phi'(\pm\infty)=0$ and $A'(y\rightarrow\pm\infty)\rightarrow{\pm}k$. Then, from Eq. (\ref{EOMThickBraneSuperpotential}), the superpotential {should satisfy} the conditions $W_{\phi}(0)=0$ and $W_{\phi}(\pm v)=0$.


\section{Conclusion and discussion}\label{conclution}

In summary, we have generalized the RS brane model as well as its smooth version in the recently proposed critical gravity theory \cite{LuPRL2011}. We found that the EOMs for the brane scenarios are of fourth order if the critical condition is not introduced, hence in this case there are no thin brane solutions~\footnote{This conclusion is obtained directly by observing the Einstein equations~\eqref{EOMThinBrane1}. In order to embed $(n-2)$-branes, one usually ask $A{''''}\sim \delta(y)$, so that $A'''$ contains a skip, while $A''$, $A'$ and $A$ are continuous. But the noncontinuous skipping function $A'''$ brings some troubles. Recall that in the RS brane model, the step function $A'\sim\epsilon(y)$ appears in the EOM in terms of $A'^2$, which is continuous. However, {in the critical gravity}, we get a noncontinuous term $A'''A'$, which cannot be canceled by other terms at the skipping point. Thus the system we considered in \eqref{actionS1}-\eqref{line-element} supports no thin brane solution if without the critical condition.}. However, {in the critical case}, the EOMs are of second order and the thin and thick brane solutions in five dimensions are obtained. All these branes are embedded in higher-dimensional AdS spacetimes.

{For the thick brane} scenario, because the scalar $\phi$ has a kink solution, the fermion zero mode can be localized on the thick branes by introducing the Yukawa coupling $\eta\bar{\Psi}\phi\Psi$ (see e.g. Refs. \cite{Volkas0705.1584,Liu:2009dw,Almeida2009}).

Brane-world {models} in higher derivative {gravity theories} were considered {for example} in Refs. \cite{NojiriOdintsov2000b,CharmousisDavisDufaux2003,AfonsoBazeiaMenezesPetrov2007,DeruelleSasakiSendouda2008,DzhunushalievFolomeevKleihausKunz2010,HoffDias2011,Zhong2011,ZhongLiuYang2011}.
Here, we compare our thick brane solutions given in this paper with the one in the $f(R)$ gravity with $f(R)=R+\alpha R^2$ \cite{Zhong2011}.
The action is
\begin{equation}\label{actionfR}
 S= \int_M  \left[
   \frac{1}{2\kappa^2}\!\left(R \!-\!3\Lambda_0 \!+\! \alpha R^2 \right)
   \!-\!\frac{1}{2}(\partial\phi)^{2}  \!-\! V(\phi)\right].~~~
\end{equation}
{The line element is the same as as (\ref{line-element}).} The thick brane is also generated by a scalar field
with the usual $\phi^4$ potential.
The EOMs are of fourth order in this $R^2$ gravity, and the solution is given by \cite{Zhong2011}
\begin{subequations}\label{{ThickBraneSolution_fR}}
\begin{eqnarray}
\phi(y)&=&v_0 \tanh(ky),\\
e^{2A(y)}&=&\cosh^{-2}(k y), \\
\Lambda_0&=&-\frac{159}{3364\alpha},
\end{eqnarray}
\end{subequations}
where $k=\sqrt{\frac{3}{232\alpha}}$.
It was shown that the linear tensor perturbation equations of the brane metric are of second order. The solution is stable against the tensor perturbations and gravity can be localized on the brane \cite{Zhong2011,ZhongLiuYang2011}. {It is still not clear whether the scalar perturbations are stable or not.}  While, for the {case} of the critical gravity, although the field equations for a brane model are of second order, {the linear tensor} perturbation equations are of fourth order.

In order to study the effective four-dimensional gravity on the branes, we need to consider the perturbations of the background metric:
\begin{eqnarray}
ds^2&=&e^{2A(z)}(\eta_{MN}+\bar{h}_{MN})dx^M dx^N.
\end{eqnarray}

In the following,  we give some arguments to simplify our calculation.

{Firstly}, in a flat spacetime, {it has been proven} that the nontransverse traceless (NT) component of the metric fluctuations just contains the following {terms}
\begin{equation}
h^{\text{NT}}_{MN}=\partial_{(M}f_{N)}(x,z)+g(x,z)\eta_{MN},
\end{equation}
for some functions $f_N (x,z)$ and $g(x,z)$. {Secondly}, the {Weyl tensor $C^M_{\;\;\;NPQ}$ is conformally invariant}, so we can calculate its perturbations in a flat spacetime. {Lastly}, since the Weyl tensor in the brane background vanishes, the tensor $\delta C^M_{~~NPQ}$ is gauge invariant. Since the NT component can be canceled by {the gauge and conformal transformations}, $\delta C^M_{\;\;\;NPQ} (\bar{h}_{RS})=\delta C^M_{\;\;\;NPQ} (\bar{h}^{\text{TT}}_{RS})$ for a flat spacetime. (Here transverse traceless (TT) means $\eta^{MP}\partial_P\bar{h}^{\text{TT}}_{MN}=0=\eta^{MN}\bar{h}^{\text{TT}}_{MN}$.)

If we choose the axial gauge $\bar{h}_{5M}=0$, then the TT condition means $\eta^{\mu\nu}\partial_\mu\bar{h}^{\text{TT}}_{\nu\rho}=0=\eta^{\mu\nu}\bar{h}^{\text{TT}}_{\mu\nu}$. Since the NT and TT components of the fluctuations {are} decoupled, and the NT components do not contribute to the $C^2$ part, the NT (scalar) perturbation equations are the same as that of the EGB gravity \cite{Giovannini:2001ta}. So, it also can be shown that the scalar perturbations are stable for our brane models, and the scalar zero modes are not localized on the brane. {This is very important for a brane model, because  localized scalar zero modes would lead to a ``fifth force" never observed and is unacceptable in the effective four-dimensional theory.}

The TT parts of the metric perturbations are governed by fourth-order differential equations at the critical point.
It is unclear whether the tensor perturbations are stable and free of ghosts, and whether the four-dimensional gravitons can be localized on the branes and the effective Newton potential can be recovered. We would like to investigate {these issues in future work}.

\section*{Acknowledgement}

The authors would like to thank the anonymous referees
whose comments largely helped us in improving the original
manuscript, and thank Prof. H. L\"u for useful discussions. YXL would like to thank Prof. J.-X. Lu for the warm hospitality during his visit to ICTS, USTC.
This work was supported by the National Natural Science Foundation of China (Grants No. 11075065 and No. 11375075),
and the Fundamental Research Funds for the Central Universities (Grants No. lzujbky-2013-18 and No. lzujbky-2013-227).  YQW was supported by the National Natural Science Foundation of China (Grants No. 11005054). SFW was supported by the National Natural Science Foundation of China (Grants No. 11275120) and China Scholarship Council. Y. Zhong was supported by the Scholarship Award for Excellent Doctoral Student granted by Ministry of Education.


\end{document}